\documentclass[preprint,amsmath,amssymb,aps,showkeys,showpacs]{revtex4}
\usepackage[english]{babel}
\usepackage{graphicx}
\usepackage{graphics}
\usepackage{amsmath}
\usepackage{dcolumn}
\usepackage{amssymb}
\usepackage{bm}
\usepackage[latin1]{inputenc}

\begin{document}
\title{A Landau-type quantization from a Lorentz symmetry violation background with crossed electric and magnetic fields }
\author{K. Bakke}
\email{kbakke@fisica.ufpb.br}
\affiliation{Departamento de F\'isica, Universidade Federal da Para\'iba, Caixa Postal 5008, 58051-970, Jo\~ao Pessoa, PB, Brazil.}

\author{H. Belich} 
\affiliation{Departamento de F\'{\i}sica e Qu\'{\i}mica, Universidade Federal do Esp\'{\i}rito Santo, Av. Fernando Ferrari, 514, Goiabeiras, 29060-900, Vit\'{o}ria, ES, Brazil.}
\email{belichjr@gmail.com}

\begin{abstract}
We investigate the arising of an analogue of the Landau quantization from a background of the violation of the Lorentz symmetry established by a time-like 4-vector and a field configuration of crossed electric and magnetic field. We also analyse the effects on this Landau-type system subject to a hard-wall confining potential by showing a particular case where a discrete spectrum of energy can be obtained. Further, we analyse the effects of a linear confining potential on the Landau-type system. We show that a quantum effect characterized by the dependence of the cyclotron frequency on the quantum numbers of the system can arise in this analogue of the Landau system. As an example, we calculate the cyclotron frequency associated with ground state of the system. 

\end{abstract}

\maketitle

\section{Introduction}

The recent discovery of the Higgs boson is finalizing a journey of incredible success of the Standard Model (SM). Nevertheless, the model depends on 19 parameters that agree with experiments, which is a signal that it is really an effective theory. However, there is the problem of the gauge hierarchy, in which the mass of the Higgs boson diverges heading towards the Planck scale. Moreover, the SM does not answer the problem of matter-antimatter asymmetry. Neutrinos appear massless in the SM, which contradicts experiments of the nineties. By analysing the luminosity of supernovas, another problem has arisen: the discovery of the universe expansion in a accelerated mode (dark energy). A previous problem of the idea of the dark energy was observed in the galaxy rotation curves, which have been much more pronounced than that expected by taking into account the luminous mass observed in the region next to the galaxies. As being the SM not intended to describe gravitation, therefore these problems are invisible to it.

Despite the successful program of SM, the SM cannot attack the points raised above, then, it is necessary to go beyond this model. Furthermore, in the SM, the unification of the fields is still incomplete. Although the electromagnetic and weak fields have been unified in the Weinberg-Salam model \cite{salam}, the strong field is apart from them (the gauge sector of the SM is $SU(3)\times SU(2)\times U(1)$, that is, QCD + electroweak). The electroweak unification is possible to be carried out thanks to the Higgs mechanism that generates mass to intermediate bosons of the weak interaction.  

A possible way of dealing with a scenario beyond the SM is the extension of the mechanism for the spontaneous symmetry breaking through vector or tensor fields, which implies that the Lorentz symmetry is violated. Kosteleck\'y and Samuelson \cite{extra3} in 1989, based on the string field theory, performed an interesting work in order that the spontaneous violation of the Lorentz symmetry can be established by non-scalar fields (vacuum of fields that have a tensor nature). On the other hand, if the spontaneous symmetry breaking is performed by a field with nonzero rank, then, it implies that the violation of the Lorentz symmetry is made by active transformations \cite{ens}. To maintain a consistent description of fluctuations around this new vacuum, it is necessary that the components of the background field to be constant. The idea behind this proposal is to investigate the presence of fields that causes the anisotropy in the spacetime, since it can affect the physical properties of the particles. The detection of fields that comes from this breaking of symmetry can be view through a change in the kinetic properties of the particles, therefore, the breaking of symmetry induces privileged directions in the spacetime that can be measured experimentally \cite{extra1, extra2, col}, and thus it can give us hints about a more fundamental theory behind the broken symmetry.

An appropriate treatment to study possible extensions of the Standard Model is to modify the Dirac theory under the influence of a Lorentz-violating background which is nonminimally coupled to a spinor \cite{ea,ea2,ea3}. Some examples that deal with the nonrelativistic limit of the modified Dirac theory are the spectrum of energy of the hydrogen atom discussed in Ref. \cite{belich2}, an analogue of the quantum Hall effect \cite{lin3} and geometric quantum phases \cite{belich,belich1,belich3,bb,bb4}.

In this paper, we follow this line of nonminimal couplings because it enables us to go beyond the scenario established by the Standard Model. Here, an analogue of the Landau quantization based on a background of the violation of the Lorentz symmetry is established by a time-like 4-vector and a field configuration of crossed electric and magnetic field is investigated. Moreover, the effects on this Landau-type system subject to a hard-wall confining potential is analysed. Finally, we analyse the effects of a linear confining potential on the Landau-type system. We show that a quantum effect characterized by the dependence of the cyclotron frequency on the quantum numbers of the system can arise in this analogue of the Landau system. As an example, we calculate the cyclotron frequency associated with ground state of the system.

The structure of this paper is: in section II, we introduce a nonminimal coupling into the Dirac equation to describe a background of the violation of the Lorentz symmetry based on fixed 4-vector and a field configuration of crossed electric and magnetic field. Then, by taking the nonrelativistic limit of the Dirac equation, we discuss a way of build an analogue of the Landau quantization based on a background established by a time-like 4-vector and a field configuration of crossed electric and magnetic field; in section III, we analyse the behaviour of the Landau-type system subject to a hard-wall confining potential; in section IV, we analyse the effects of a linear confining potential on the Landau-type system; in section V, we present ours conclusions.

\section{Landau-type quantization}

Recently, we have proposed to introduce a new term into the Dirac equation with the purpose of obtaining a Rashba-type coupling and a geometric quantum phase in the nonrelativistic limit of the Dirac equation that stem from a background of the violation of the Lorentz symmetry established by a time-like 4-vector and a field configuration of crossed electric and magnetic field \cite{bb6}. This new term consists in a nonminimal coupling given by ($\hbar=c=1$) 
\begin{eqnarray}
i\gamma^{\mu}\partial_{\mu}\rightarrow i\gamma^{\mu}\partial_{\mu}-\frac{g}{2}\,\eta^{\alpha\beta}\,\bar{F}_{\mu\alpha}\left(x\right)\,\bar{F}_{\beta\nu}\left(x\right)\,\gamma^{\mu}\,b_{\lambda}\,\gamma^{\lambda}\,\gamma^{\nu},
\label{1.1}
\end{eqnarray}
where $g$ is a constant, $b^{\mu}$ is a fixed $4$-vector that acts as a vector field which breaks the Lorentz symmetry and the tensor $\bar{F}_{\mu\nu}\left(x\right)$is the dual electromagnetic tensor, whose components are $\bar{F}_{0i}=-\bar{F}_{i0}=B_{i}$ and $\bar{F}_{ij}=-\bar{F}_{ji}=\epsilon_{ijk}E^{k}$. The $\gamma^{\mu}$ matrices are defined in the Minkowski spacetime in the form \cite{greiner}:
\begin{eqnarray}
\gamma^{0}=\hat{\beta}=\left(
\begin{array}{cc}
1 & 0 \\
0 & -1 \\
\end{array}\right);\,\,\,\,\,\,
\gamma^{i}=\hat{\beta}\,\hat{\alpha}^{i}=\left(
\begin{array}{cc}
 0 & \sigma^{i} \\
-\sigma^{i} & 0 \\
\end{array}\right);\,\,\,\,\,\,\Sigma^{i}=\left(
\begin{array}{cc}
\sigma^{i} & 0 \\
0 & \sigma^{i} \\	
\end{array}\right),
\label{1.2}
\end{eqnarray}
with $\vec{\Sigma}$ as being the spin vector. The matrices $\sigma^{i}$ are the Pauli matrices and satisfy the relation $\left(\sigma^{i}\,\sigma^{j}+\sigma^{j}\,\sigma^{i}\right)=2\eta^{ij}$. 

In what follows, we wish to work with the cylindrical symmetry, therefore we need to introduce the covariant form of the Dirac equation in order to deal with curvilinear coordinates. By following Ref. \cite{schu}, we need to apply a coordinate transformation $\frac{\partial}{\partial x^{\mu}}=\frac{\partial \bar{x}^{\nu}}{\partial x^{\mu}}\,\frac{\partial}{\partial\bar{x}^{\nu}}$ and a unitary transformation on the wave function $\Psi\left(x\right)=U\,\Psi'\left(\bar{x}\right)$ in order that the Dirac equation can be written in any orthogonal system. Hence, the term $i\gamma^{\mu}\partial_{\mu}$ of the Dirac equation must be written as $i\gamma^{\mu}\partial_{\mu}\rightarrow i\,\gamma^{\mu}\,D_{\mu}\,+\frac{i}{2}\,\sum_{k=1}^{3}\,\gamma^{k}\,\left[D_{k}\,\ln\left(\frac{h_{1}\,h_{2}\,h_{3}}{h_{k}}\right)\right]$, where $D_{\mu}=\frac{1}{h_{\mu}}\,\partial_{\mu}$ is the derivative of the corresponding coordinate system and the parameter $h_{k}$ corresponds to the scale factors of this coordinate system. For example, let us take the line element of the Minkowski spacetime in cylindrical coordinates ($\hbar=c=1$): $ds^{2}=-dt^{2}+d\rho^{2}+\rho^{2}d\varphi^{2}+dz^{2}$; thus, the scale factors are $h_{0}=1$, $h_{1}=1$, $h_{2}=\rho$ and $h_{3}=1$. In this way, by introducing the nonminimal coupling (\ref{1.1}), the Dirac equation becomes 
\begin{eqnarray}
i\,\gamma^{\mu}\,D_{\mu}\,\Psi+\frac{i}{2}\,\sum_{k=1}^{3}\,\gamma^{k}\,\left[D_{k}\,\ln\left(\frac{h_{1}\,h_{2}\,h_{3}}{h_{k}}\right)\right]\Psi-\frac{g}{2}\,\eta^{\alpha\beta}\,\bar{F}_{\mu\alpha}\left(x\right)\,\bar{F}_{\beta\nu}\left(x\right)\,\gamma^{\mu}\,b_{\lambda}\,\gamma^{\lambda}\,\gamma^{\nu}\Psi=m\Psi.
\label{1.3}
\end{eqnarray}
It is worth mentioning that the second term of the Dirac equation (\ref{1.3}) corresponds to a term called as the spinorial connection in which is introduced in the Dirac equation when we are dealing with the quantum field theory in curved space \cite{bd}.

Our focus is on a background of the Lorentz symmetry violation determined by a time-like 4-vector $b_{\lambda}=\left(b_{0},0,0,0\right)$, then, the Dirac equation (\ref{1.3}) can be written in the form:
\begin{eqnarray}
i\frac{\partial\Psi}{\partial t}=m\hat{\beta}\Psi+\vec{\alpha}\cdot\vec{\pi}\,\Psi+g\,b_{0}\vec{\alpha}\cdot\left(\vec{E}\times\vec{B}\right)\Psi-gb_{0}E^{2}\,\Psi,
\label{1.4}
\end{eqnarray}
where we have defined the operator $\vec{\pi}=\vec{p}-i\vec{\xi}$, whose vector $\vec{\xi}$ is given in such a way that its components are $-i\xi_{k}=-\frac{\sigma^{3}}{2\rho}\,\delta_{2k}$ \cite{bb6}.

In this section, we wish to discuss the nonrelativistic quantum dynamics of a Dirac neutral particle in the background established by the time-like vector $b_{\lambda}=\left(b_{0},0,0,0\right)$ and a field configuration of crossed electric and magnetic fields. For this reason, let us we apply the Foldy-Wouthuysen approximation \cite{fw,greiner} up to the terms of order $m^{-1}$. In this approach, we need first to write the Dirac equation in the form:
\begin{eqnarray}
i\,\frac{\partial\Psi}{\partial t}=\hat{H}\,\Psi,
\label{1.5}
\end{eqnarray} 
where the Hamiltonian operator $\hat{H}$ of the system must be written in terms of even operator $\hat{\epsilon}$ and odd operators $\hat{\mathcal{O}}$ as: $\hat{H}=m\,\hat{\beta}+\hat{\mathcal{E}}+\hat{\mathcal{O}}$. Both even and odd operators must be Hermitian operators and satisfy the following relations: $\left[\hat{\mathcal{E}},\hat{\beta}\right]=\hat{\mathcal{E}}\,\hat{\beta}-\hat{\beta}\,\hat{\mathcal{E}}=0$ and $\left\{\hat{\mathcal{O}},\hat{\beta}\right\}=\hat{\mathcal{O}}\,\hat{\beta}+\hat{\beta}\,\hat{\mathcal{O}}=0$. In short, the objective of the Foldy-Wouthuysen approach \cite{fw,greiner} is to apply a unitary transformation in order to remove the operators from the Dirac equation that couple the ``large'' to the ``small'' components of the Dirac spinors. We have that the even operators $\hat{\mathcal{E}}$ do not couple the ``large'' to the ``small'' components of the Dirac spinors, while the odd operators $\hat{\mathcal{O}}$ do couple them. Thereby, by applying he Foldy-Wouthuysen approximation \cite{fw,greiner} up to the terms of order $m^{-1}$, we can write the nonrelativistic limit of the Dirac equation in the form:
\begin{eqnarray}
i\frac{\partial\Psi}{\partial t}=m\hat{\beta}\Psi+\hat{\mathcal{E}}\Psi+\frac{\hat{\beta}}{2m}\,\hat{\mathcal{O}}^{2}\Psi
\label{1.6}
\end{eqnarray}

Hence, from Eq. (\ref{1.4}), the operators $\hat{\mathcal{O}}$ and $\hat{\mathcal{E}}$ are
\begin{eqnarray}
\hat{\mathcal{O}}=\vec{\alpha}\cdot\vec{\pi}+g\,b_{0}\,\vec{\alpha}\cdot\left(\vec{E}\times\vec{B}\right);\,\,\,\,\,\,\,\,\,\hat{\mathcal{E}}=-g\,b_{0}\,E^{2}.
\label{1.7}
\end{eqnarray}

By substituting (\ref{1.7}) into (\ref{1.6}), we obtain the Schr\"odinger equation (for 2-spinors):
\begin{eqnarray}
i\frac{\partial\psi}{\partial t}=\frac{1}{2m}\left[\vec{p}+gb_{0}\left(\vec{E}\times\vec{B}\right)\right]^{2}\psi+\frac{1}{2m}\,gb_{0}\,\vec{\Sigma}\cdot\vec{B}_{\mathrm{eff}}\,\psi-gb_{0}\,E^{2}\,\psi.
\label{1.8}
\end{eqnarray}

Observe that we can defined a connection 1-form $A_{\mu}^{\mathrm{eff}}\left(x\right)$ from Eq. (\ref{1.8}) in such a way that the time-like component $A_{0}^{\mathrm{eff}}\left(x\right)$ corresponds to an effective scalar potential and space-like component $\vec{A}_{\mathrm{eff}}\left(x\right)$ corresponds to an effective potential vector, that is,
\begin{eqnarray}
A_{0}^{\mathrm{eff}}\left(x\right)=E^{2};\,\,\,\,\vec{A}_{\mathrm{eff}}\left(x\right)=\vec{E}\times\vec{B}.
\label{1.9}
\end{eqnarray}
Besides, we can define effective magnetic electric fields as
\begin{eqnarray}
\vec{B}_{\mathrm{eff}}&=&\vec{\nabla}\times\vec{A}_{\mathrm{eff}}=\vec{\nabla}\times\left(\vec{E}\times\vec{B}\right);\nonumber\\
[-2mm]\label{1.10}\\[-2mm]
\vec{E}_{\mathrm{eff}}&=&-\vec{\nabla}A_{0}^{\mathrm{eff}}=-\vec{\nabla}E^{2}.\nonumber
\end{eqnarray}

Based on the system of a neutral particle with an induced electric dipole moment that interacts with electric and magnetic fields, Furtado {\it el al} \cite{lin2} proposed an analogue of the Landau quantization. In analogy with Ref. \cite{lin2}, we have proposed recently a Landau-type quantization by assuming that the fixed vector background is a time-like vector $b_{\lambda}=\left(b_{0},0,0,0\right)$ and electric and magnetic fields must be determined in a such a way that the effective magnetic field given in Eq. (\ref{1.10}) is uniform \cite{bb6}. For this purpose, we have used the following field configuration \cite{lin2}: $\vec{E}=\frac{\lambda\,\rho}{2}\,\hat{\rho}$ and $\vec{B}=B_{0}\,\hat{z}$, where $\lambda$ is a constant associated with a uniform volume electric charge density, $\rho=\sqrt{x^{2}+y^{2}}$, $\hat{\rho}$ is a unit vector in the radial direction, $\hat{z}$ is the unit vector in the $z$ direction and $B_{0}$ is a constant. With this field configuration, then, we have that the effective magnetic field defined in Eq. (\ref{1.10}) is uniform and we have a Landau-type system in a background of the violation of the Lorentz symmetry. However, due to the presence of the term proportional to $E^{2}$ in Eq. (\ref{1.8}), the field configuration above creates a problem in determining the cyclotron frequency of the system, because it can be or not a real number. Unless we introduce a scalar potential, as we did in Ref. \cite{bb6} by confining the neutral particle to a two-dimensional quantum ring, hence, this field configuration is not the best choice of building a Landau-like system. 

On the other hand, a possible field configuration is
\begin{eqnarray}
\vec{B}=\frac{\lambda_{m}\,\rho}{2}\,\hat{\rho};\,\,\,\,\vec{E}=E_{0}\,\hat{z},
\label{1.12}
\end{eqnarray} 
where $\lambda_{m}$ is a constant related to a uniform volume density of magnetic charges and $E_{0}$ is a constant. The magnetic field given in Eq. (\ref{1.12}) was proposed in Ref. \cite{lin} with the aim of establish the analogue of the Landau quantization for a neutral particle with a permanent electric dipole moment. Note that this field configuration (\ref{1.12}) gives rise to a uniform effective magnetic field in Eq. (\ref{1.10}) and a null effective electric field given by: 
\begin{eqnarray}
\vec{B}_{\mathrm{eff}}=\lambda_{m}\,E_{0}\,\hat{z};\,\,\,\,\vec{E}_{\mathrm{eff}}=0;
\label{1.13}
\end{eqnarray}
thus the field configuration (\ref{1.12}) establishes a Landau-type system in a background of the violation of the Lorentz symmetry without any problem in determining the cyclotron frequency due to the presence of the term proportional to $E^{2}$ in Eq. (\ref{1.8}). In this way, the Schr\"odinger equation (\ref{1.8}) becomes
\begin{eqnarray}
i\frac{\partial\psi}{\partial t}&=&-\frac{1}{2m}\left[\frac{\partial^{2}}{\partial\rho^{2}}+\frac{1}{\rho}\frac{\partial}{\partial\rho}+\frac{1}{\rho^{2}}\frac{\partial^{2}}{\partial\varphi^{2}}+\frac{\partial^{2}}{\partial z^{2}}\right]\psi+\frac{1}{2m}\frac{i\sigma^{3}}{\rho^{2}}\frac{\partial\psi}{\partial\varphi}+\frac{1}{8m\rho^{2}}\psi\nonumber\\
[-2mm]\label{1.14}\\[-2mm]
&-&i\frac{gb_{0}\lambda_{m} E_{0}}{2m}\frac{\partial\psi}{\partial\varphi}+\frac{gb_{0}\lambda_{m} E_{0}}{4m}\,\sigma^{3}\,\psi+\frac{\left(gb_{0}\lambda_{m} E_{0}\right)^{2}}{8m}\,\rho^{2}\,\psi-gb_{0}E_{0}^{2}\psi.\nonumber
\end{eqnarray}

Note that $\psi$ is an eigenfunction of $\sigma^{3}$ in Eq. (\ref{1.14}), whose eigenvalues are $s=\pm1$. Thereby, we can write $\sigma^{3}\psi_{s}=\pm\psi_{s}=s\psi_{s}$. Moreover, note that the operators $\hat{p}_{z}=-i\partial_{z}$ and $\hat{J}_{z}=-i\partial_{\varphi}$ \cite{schu} commute with the Hamiltonian of the right-hand side of (\ref{1.14}), therefore, a particular solution to Eq. (\ref{1.14}) can be written in terms of the eigenvalues of the operator $\hat{p}_{z}$ and $\hat{J}_{z}$ \footnote{It has been shown in Ref. \cite{schu} that the $z$-component of the total angular momentum in cylindrical coordinates is given by $\hat{J}_{z}=-i\partial_{\varphi}$, where the eigenvalues are $\mu=l\pm\frac{1}{2}$.}: 
\begin{eqnarray}
\psi_{s}=e^{-i\mathcal{E}t}\,e^{i\left(l+\frac{1}{2}\right)\varphi}\,e^{ikz}\,R_{s}\left(\rho\right),
\label{1.15}
\end{eqnarray}
where $l=0,\pm1,\pm2,\ldots$ and $k$ is a constant. From now on, we consider $k=0$ in order to describe a planar system. Substituting the solution (\ref{1.15}) into the Schr\"odinger equation (\ref{1.14}), we obtain the following radial equation:
\begin{eqnarray}
R_{s}''+\frac{1}{\rho}R_{s}'-\frac{\gamma_{s}^{2}}{\rho^{2}}R_{s}-\frac{\left(gb_{0}\lambda_{m} E_{0}\delta\right)^{2}}{4}\,\rho^{2}\,R_{s}+\zeta\,R_{s}=0,
\label{1.16}
\end{eqnarray}
where we have defined in Eq. (\ref{1.16}) the parameters:
\begin{eqnarray}
\gamma_{s}&=&l+\frac{1}{2}\left(1-s\right)\nonumber\\
[-2mm]\label{1.17}\\[-2mm]
\zeta&=&2m\mathcal{E}-gb_{0}\lambda_{m} E_{0}\,\gamma_{s}-s\,gb_{0}\lambda_{m} E_{0}+gb_{0}E_{0}^{2}.\nonumber
\end{eqnarray}

Let us perform a change of variables given by $\xi=\frac{gb_{0}\lambda_{m} E_{0}}{2}\,\rho^{2}$. In this way, the radial equation (\ref{1.16}) becomes
\begin{eqnarray}
\xi\,R_{s}''+R_{s}'-\frac{\gamma_{s}^{2}}{4\xi}R_{s}-\frac{\xi}{4}\,R_{s}+\frac{\zeta}{2gb_{0}\lambda_{m} E_{0}}\,R_{s}=0.
\label{1.18}
\end{eqnarray}

In order that a regular solution at the origin can be obtained, the solution to the second order differential equation (\ref{1.18}) can be given in the form:
\begin{eqnarray}
R_{s}\left(\xi\right)=e^{-\frac{\xi}{2}}\,\xi^{\frac{\left|\tau\right|}{2}}\,M_{s}\left(\xi\right),
\label{1.19}
\end{eqnarray}
where $M_{s}\left(\xi\right)$ is an unknown function. Substituting (\ref{1.19}) into (\ref{1.18}), we obtain the following second-order differential equation:
\begin{eqnarray}
\xi\,M_{s}''+\left[\left|\gamma_{s}\right|+1-\xi\right]M_{s}'+\left[\frac{\zeta}{2gb_{0}\lambda_{m} E_{0}\delta}-\frac{\left|\gamma_{s}\right|}{2}-\frac{1}{2}\right]M_{s}=0.
\label{1.20}
\end{eqnarray}

Equation (\ref{1.20}) is the Kummer equation or the confluent hypergeometric function \cite{abra}. In order to obtain a solution to Eq. (\ref{1.20}) which is regular at the origin, we consider only the Kummer function of first kind given by $M_{s}\left(\xi\right)=M\left(\frac{\left|\gamma_{s}\right|}{2}+\frac{1}{2}-\frac{\zeta}{2gb_{0}\lambda_{m} E_{0}},\,\left|\gamma_{s}\right|+1,\,\xi\right)$ \cite{abra}. Moreover, a normalized radial wave function can be obtained if we impose that the hypergeometric series becomes a polynomial of degree $n$ ($n=0,1,2,\ldots$). This makes the radial wave function to be finite everywhere \cite{landau}. Hence, a finite radial solution to Eq. (\ref{1.20}) can be achieved when the parameter $\frac{\left|\tau\right|}{2}+\frac{1}{2}-\frac{\zeta}{2gb_{0}\lambda_{m} E_{0}}$ of the Kummer function is equal to a non-positive integer number, that is, when $\frac{\left|\gamma_{s}\right|}{2}+\frac{1}{2}-\frac{\zeta}{2gb_{0}\lambda_{m} E_{0}}=-n$ ($n=0,1,2,\ldots$). With this condition, the energy levels of the bound states are
\begin{eqnarray}
\mathcal{E}_{n,\,l,\,s}=\omega\left[n+\frac{\left|\gamma_{s}\right|}{2}+\frac{\gamma_{s}}{2}+\frac{1}{2}+\frac{s}{2}\right]-gb_{0}E_{0}^{2},
\label{1.21}
\end{eqnarray}
where the corresponding to the angular frequency or the cyclotron frequency is 
\begin{eqnarray}
\omega=\frac{gb_{0}\lambda_{m} E_{0}}{m}.
\label{1.21a}
\end{eqnarray} 

The energy levels (\ref{1.21}) correspond to the analogue of the Landau levels for a spin-1/2 neutral particle based on the Lorentz symmetry violation background established by a time-like fixed vector and a field configuration of crossed electric and magnetic fields (\ref{1.12}).

\section{Landau-type system subject to a hard-wall confining potential}

In this section, we discuss the confinement of the Landau-type system that stems from the Lorentz symmetry violation background established by a time-like fixed vector and the field configuration given in Eq. (\ref{1.12}) to a hard-wall confining potential. For this purpose, let us also assume that the wave function is well-behaved at the origin as in the previous section, and thus assume that the wave function vanishes at a fixed radius $\rho_{0}$ \cite{fur,val2,bf,bf2}, that is,
\begin{eqnarray}
R\left(\xi_{0}=\frac{gb_{0}\lambda_{m} E_{0}^{2}}{2}\,\rho^{2}_{0}\right)=0.
\label{2.1}
\end{eqnarray}

As we have seen in the previous section, the radial coordinate is defined in a two-dimensional system in the range $0\,<\,\rho\,<\,\infty$, however, if this system is also subject to a hard-wall confining potential, then, the radial coordinate becomes defined in the range $0\,<\,\rho\,<\,\rho_{0}$, where $\rho_{0}$ is finite ($\rho_{0}<\infty$). Therefore, we cannot impose anymore that the confluent hypergeometric series becomes a polynomial of degree $n$, since we wish to obtain a normalized wave function in the range $0\,<\,\rho\,<\,\rho_{0}$. Henceforth, let us discuss a particular case where a discrete spectrum of energy can be obtained. First of all, note that the radial wave function (\ref{1.19}) is written in form:
\begin{eqnarray}
R_{s}\left(\xi\right)=e^{-\frac{\xi}{2}}\,\xi^{\frac{\left|\tau\right|}{2}}\,M\left(\frac{\left|\gamma_{s}\right|}{2}+\frac{1}{2}-\frac{\zeta}{2gb_{0}\lambda_{m} E_{0}},\,\left|\gamma_{s}\right|+1,\,\xi\right).
\label{2.2}
\end{eqnarray}

Based on Refs. \cite{abra}, we can take a fixed value for the parameter $B=\left|\gamma_{s}\right|+1$ of the confluent hypergeometric function and, as a consequence, we can consider the parameter $A=\frac{\left|\gamma_{s}\right|}{2}+\frac{1}{2}-\frac{\zeta}{2gb_{0}\lambda_{m} E_{0}}$ of the confluent hypergeometric function to be large because the product $gb_{0}$ is considered to be small. Therefore, for a fixed radius $\rho_{0}$, the confluent hypergeometric function can be written as follows \cite{abra}: 
\begin{eqnarray}
M\left(A,B,\xi_{0}=\frac{gb_{0}\lambda_{m} E_{0}}{2}\,\rho^{2}_{0}\right)&\approx&\frac{\Gamma\left(B\right)}{\sqrt{\pi}}\,e^{\frac{\xi_{0}}{2}}\left(\frac{B\xi_{0}}{2}-A\xi_{0}\right)^{\frac{1-B}{2}}\times\nonumber\\
[-2mm]\label{2.3}\\[-2mm]
&\times&\cos\left(\sqrt{2B\xi_{0}-4A\xi_{0}}-\frac{B\pi}{2}+\frac{\pi}{4}\right),\nonumber
\end{eqnarray}
where $\Gamma\left(B\right)$ is the gamma function. Next, by applying the boundary condition established in Eq. (\ref{2.1}), we obtain the following expression for the energy levels:
\begin{eqnarray}
\mathcal{E}_{n,\,l}\approx\frac{1}{2m\rho_{0}^{2}}\left[n\pi+\frac{\left|\gamma_{s}\right|}{2}\pi+\frac{3\pi}{4}\right]^{2}+\omega\left[\frac{\gamma_{s}}{2}+\frac{s}{2}\right]-gb_{0}E_{0}^{2}.
\label{2.4}
\end{eqnarray}

Despite the presence of an effective uniform magnetic field (\ref{1.13}), the energy levels (\ref{2.4}) correspond to the spectrum of energy of a Landau-type system subject to a hard-wall confining potential \cite{bf,bf2}. Observe that the energy levels (\ref{2.4})) are parabolic with respect to the quantum number $n$ in contrast to the analogue of the Landau levels in Eq. (\ref{1.21}) in which are nonparabolic with respect to the quantum number $n$. The discrete spectrum of energy (\ref{2.4}) is obtained due to the assumption that the parameters of the violation of the Lorentz symmetry are small, that is, the product $gb_{0}$ is small. If this product would not be small, then, the discrete spectrum of energy (\ref{2.4}) could not be determined.

\section{Landau-type system subject to a linear confining potential}

In recent decades, a discussion about the way of introducing a scalar potential (non-electromagnetic potential) in a relativistic equation could be through a modification of the mass term \cite{greiner,scalar} has been performed. The well-known procedure is to consider an electric charged particle that interacts with the electromagnetic field and introduce a minimal coupling into the Dirac and Klein-Gordon equations, that is, a scalar potential can be introduced by modifying the momentum operator $p_{\mu}=i\partial_{\mu}$ via relation $p_{\mu}\rightarrow p_{\mu}-q\,A_{\mu}\left(x\right)$. On the other hand, in Ref. \cite{scalar} is shown that a scalar potential (non-electromagnetic potential) can be introduced by modifying the mass term as $m\rightarrow m+S\left(\vec{r},\,t\right)$, where $S\left(\vec{r},\,t\right)$ is the scalar potential. This modification of the mass term has been explored in several topics, such as the scalar field in the cosmic string spacetime \cite{eug}, a Dirac particle in the presence of static scalar potential, a Coulomb potential \cite{scalar2}, the quark-antiquark interaction as a problem of a relativistic spin-$0$ possessing a position-dependent mass \cite{bah} and with the Klein-Gordon oscillator \cite{bf3}. A particular interest in introducing a scalar potential as an additional term of the mass of the particle is in studies of confinement of quarks, such as in the {\it bag} model, since it can generalize the rest mass of the particle to the model of confinement of quarks \cite{quark,quark2}. This confinement of quarks is described by a linear confining potential. 

In this section, let us introduce a linear scalar potential into the Dirac equation as a modification of the mass term and investigate its influence on the Landau-type system established by the Lorentz symmetry violation background established by a time-like fixed vector and the field configuration given in Eq. (\ref{1.12}). Let us consider the linear confining potential given by
\begin{eqnarray}
S\left(\rho\right)=\eta\,\rho,
\label{3.1}
\end{eqnarray}
where $\eta$ is a parameter that characterizes the scalar potential. By following the Foldy-Wouthuysen approximation \cite{fw,greiner}, we should note that the even operator given in Eq. (\ref{1.7}) is thus given by $\hat{\mathcal{E}}=-g\,b_{0}\,E^{2}+\hat{\beta}\,S\left(\rho\right)$. Thereby, the Schr\"odinger equation (\ref{1.14}) becomes
\begin{eqnarray}
i\frac{\partial\psi}{\partial t}&=&-\frac{1}{2m}\left[\frac{\partial^{2}}{\partial\rho^{2}}+\frac{1}{\rho}\frac{\partial}{\partial\rho}+\frac{1}{\rho^{2}}\frac{\partial^{2}}{\partial\varphi^{2}}+\frac{\partial^{2}}{\partial z^{2}}\right]\psi+\frac{1}{2m}\frac{i\sigma^{3}}{\rho^{2}}\frac{\partial\psi}{\partial\varphi}+\frac{1}{8m\rho^{2}}\psi-i\frac{gb_{0}\lambda_{m} E_{0}}{2m}\frac{\partial\psi}{\partial\varphi}\nonumber\\
[-2mm]\label{3.1a}\\[-2mm]
&+&\frac{gb_{0}\lambda_{m} E_{0}}{4m}\,\sigma^{3}\,\psi+\frac{\left(gb_{0}\lambda_{m} E_{0}\right)^{2}}{8m}\,\rho^{2}\,\psi-gb_{0}E_{0}^{2}\psi+S\left(\rho\right)\psi.\nonumber
\end{eqnarray}

By following the steps from Eq. (\ref{1.15}) to Eq. (\ref{1.17}), then, the Schr\"odinger equation (\ref{3.1a}) becomes  
\begin{eqnarray}
R_{s}''+\frac{1}{\rho}R_{s}'-\frac{\gamma_{s}^{2}}{\rho^{2}}R_{s}-\frac{\left(gb_{0}\lambda_{m} E_{0}\delta\right)^{2}}{4}\,\rho^{2}\,R_{s}-2m\eta\rho\,R_{s}+\zeta\,R_{s}=0.
\label{3.2}
\end{eqnarray}

Let us perform a new change of variables given by $r=\sqrt{\frac{gb_{0}\lambda_{m}E_{0}}{2}}\,\rho$, then, Eq. (\ref{3.2}) becomes
\begin{eqnarray}
R_{s}''+\frac{1}{r}R_{s}'-\frac{\gamma_{s}^{2}}{r^{2}}R_{s}-r^{2}\,R_{s}-\delta\rho\,R_{s}+\bar{\zeta}\,R_{s}=0,
\label{3.3}
\end{eqnarray}
where we have defined two new parameters as
\begin{eqnarray}
\delta=2m\eta\,\left(\frac{2}{gb_{0}\lambda_{m}E_{0}}\right)^{3/2};\,\,\,\bar{\zeta}=\frac{2\zeta}{gb_{0}\lambda_{m}E_{0}}.
\label{3.4}
\end{eqnarray}

The asymptotic behaviour of the possible solutions to Eq. (\ref{3.3}) is determined for $r\rightarrow0$ and $r\rightarrow\infty$. Based on Refs. \cite{eug,vercin,mhv,bf3}, the behaviour of the possible solutions to Eq. (\ref{3.3}) at $r\rightarrow0$ and $r\rightarrow\infty$ allows us to write the function $R_{s}\left(r\right)$ in terms of an unknown function $H_{s}\left(r\right)$ as
\begin{eqnarray}
R_{s}\left(r\right)=e^{-r^{2}/2}\,e^{-\delta r/2}\,r^{\left|\gamma_{s}\right|}\,H_{s}\left(r\right).
\label{3.5}
\end{eqnarray}
Substituting Eq. (\ref{3.5}) into Eq. (\ref{3.3}) we have
\begin{eqnarray}
H_{s}''+\left[\frac{\theta}{r}-\delta-2r\right]H_{s}'+\left[g-\frac{\delta\theta}{2r}\right]H_{s}=0,
\label{3.6}
\end{eqnarray}
where the parameters $\theta$ and $g$ are defined as
\begin{eqnarray}
\theta=2\left|\gamma_{s}\right|+1;\,\,\,\,\,g=\bar{\zeta}-2-2\left|\gamma_{s}\right|+\frac{\delta^{2}}{4}.
\label{3.7}
\end{eqnarray}

The second order differential equation (\ref{3.6}) is known in the literature as the biconfluent Heun equation \cite{heun,eug,bf3}, and thus the function $H_{s}\left(r\right)$ is the biconfluent Heun function:
\begin{eqnarray}
H_{s}\left(r\right)=H\left(2\left|\gamma_{s}\right|,\,\delta,\,\bar{\zeta}+\frac{\delta^{2}}{4},\,0,\,r\right).
\label{3.8}
\end{eqnarray}

Let us proceed with our discussion by using the Frobenius method \cite{arf,eug}. This method permit us to write the solutions to Eq. (\ref{3.7}) as a power series expansion around the origin:
\begin{eqnarray}
H_{s}\left(r\right)=\sum_{k=0}^{\infty}d_{k}\,r^{k}.
\label{3.9}
\end{eqnarray}

After substituting Eq. (\ref{3.9}) into Eq. (\ref{3.6}), we obtain a recurrence relation given by
\begin{eqnarray}
d_{k+2}=\frac{\delta}{2}\frac{\left[2\left(k+1\right)+\theta\right]}{\left(k+2\right)\left(k+1+\theta\right)}\,d_{k+1}-\frac{\left[g-2k\right]}{\left(k+2\right)\left(k+1+\theta\right)}\,d_{k}.
\label{3.10}
\end{eqnarray}

By starting with $d_{0}=1$, then, we can calculate other coefficients of the power series expansion (\ref{3.9}). For example, the coeficients $d_{1}$ and $d_{2}$ are 
\begin{eqnarray}
d_{1}=\frac{\delta}{2};\,\,\,\,\,\,d_{2}=\frac{\delta^{2}}{8}\frac{\left(\theta+2\right)}{\left(\theta+1\right)}-\frac{g}{2\left(\theta+1\right)}.
\label{3.11}
\end{eqnarray}

Our focus in this section is on the bound states solutions to the Schr\"odinger equation (\ref{3.1a}). Bound state solutions can be achieved by imposing that the power series expansion (\ref{3.9}) becomes a polynomial of degree $n$. From the recurrence relation (\ref{3.10}), the biconfluent Heun series (\ref{3.9}) becomes a polynomial of degree $n$ by imposing the following conditions \cite{heun,eug,bf3}:
\begin{eqnarray}
g=2n;\,\,\,\,d_{n+1}=0
\label{3.12}
\end{eqnarray}
where $n=1,2,3,\ldots$. From the condition $g=2n$ given in Eq. (\ref{3.12}), we obtain the energy levels:
\begin{eqnarray}
\mathcal{E}_{n,\,l,\,s}=\frac{\omega}{2}\left[n+\left|\gamma_{s}\right|+1\right]-\frac{2\eta^{2}}{m\,\omega^{2}}+\omega\left[\frac{\gamma_{s}}{2}+\frac{s}{2}\right]-gb_{0}E_{0}^{2},
\label{3.13}
\end{eqnarray}
where $\omega$ is the cyclotron frequency given in Eq. (\ref{1.21a}).

On the other hand, we need to analyse the condition $d_{n+1}=0$ given in Eq. (\ref{3.12}) yet. For this purpose, let us assume that the electric field can be adjusted in order that the condition $d_{n+1}=0$ can be satisfied, therefore we can adjust the angular frequency $\omega$ in such a way that the condition $d_{n+1}=0$ is satisfied. This means that not all values of the angular frequency $\omega$ are allowed in the system in order that a polynomial solution to Eq. (\ref{3.6}) can be obtained. Only some specific values of $\omega$ are allowed and depend on the quantum numbers of the systems $\left\{n,\,l,\,s\right\}$. For this reason, we label $E_{0}=E_{0}^{n,\,l,\,s}$ and $\omega=\omega_{n,\,l,\,s}$ \cite{bf3}. From this assumptions, we have that the conditions given in Eq. (\ref{3.12}) are satisfied and a polynomial solution to the function $H_{s}\left(r\right)$ in Eq. (\ref{3.9}) is achieved. 

As an example, let us consider the ground state $n=1$ and analyse the condition $d_{n+1}=0$. For $n=1$, we have an angular frequency given by
\begin{eqnarray}
\omega_{1,\,l,\,s}=\left[\frac{4\eta^{2}}{m}\left(\theta+2\right)\right]^{1/3},
\label{3.13}
\end{eqnarray}
and the expression for the energy level of the ground state is
\begin{eqnarray}
\mathcal{E}_{1,\,l,\,s}=\frac{\omega_{1,\,l,\,s}}{2}\,\left[\left|\gamma_{s}\right|+2\right]-\frac{2\eta^{2}}{m\,\omega^{2}_{1,\,l,\,s}}+\omega_{1,\,l,\,s}\left[\frac{\gamma_{s}}{2}+\frac{s}{2}\right]-gb_{0}\left(E_{0}^{1,\,l,\,s}\right)^{2},
\label{3.14}
\end{eqnarray}
Hence, the radial wave function (\ref{3.5}) associated with the ground state is
\begin{eqnarray}
R_{s}\left(r\right)=e^{-r^{2}/2}\,e^{-\delta r/2}\,r^{\left|\gamma_{s}\right|}\,\left(1+\frac{\delta}{2}\,r\right).
\label{3.15}
\end{eqnarray}

Finally, the general expression of the energy levels (\ref{3.13}) must be rewritten in the form:
\begin{eqnarray}
\mathcal{E}_{n,\,l,\,s}=\frac{\omega_{n,\,l,\,s}}{2}\left[n+\left|\gamma_{s}\right|+1\right]-\frac{2\eta^{2}}{m\,\omega^{2}_{n,\,l,\,s}}+\omega_{n,\,l,\,s}\left[\frac{\gamma_{s}}{2}+\frac{s}{2}\right]-gb_{0}\left(E_{0}^{n,\,l,\,s}\right)^{2}.
\label{3.16}
\end{eqnarray}

Hence, the influence of a linear scalar potential on the Landau-type system that stems from Lorentz symmetry breaking effects gives rise to a different spectrum of energy, where the ground state is defined by the quantum number $n=1$ instead of the quantum number $n=0$ obtained in Eq. (\ref{1.21}). Besides, a quantum effect can arises in the Landau-like system characterized by the dependence of the cyclotron frequency $\omega$ on the quantum numbers $\left\{n,\,l,\,s\right\}$ of the system, whose meaning is that not all values of the cyclotron frequency are allowed in order that a polynomial solution to the function $H_{s}\left(r\right)$ given in Eq. (\ref{3.9}) can be obtained, and thus bound states can be achieved.

\section{conclusions}

In this work, we have investigated the arising of an analogue of the Landau quantization from a background of the violation of the Lorentz symmetry established by a time-like 4-vector and a field configuration of crossed electric and magnetic field. In order to achieve the analogue of the Landau quantization, we have proposed a field configuration given by a uniform electric field perpendicular to the plane of motion of the particle and a radial magnetic field produced by a uniform volume density of magnetic charges. 

We have also discussed the behaviour of the Landau-type system subject to a hard-wall confining potential. We have seen that a discrete spectrum of energy can be obtained by assuming that the parameters associated with the violation of the Lorentz symmetry are small, otherwise this discrete spectrum of energy (\ref{2.4})cannot be determined. In this case, the energy levels (\ref{2.4}) are parabolic with respect to the quantum number $n$ in contrast to the analogue of the Landau levels in Eq. (\ref{1.21}) in which are linear with respect to the quantum number $n$. 

Finally, we have analysed the influence of a linear scalar potential on the Landau-type system. We have obtained a spectrum of energy whose ground state is defined by the quantum number $n=1$ instead of the quantum number $n=0$ associated with the analogue of the Landau levels. Furthermore, a quantum effect characterized by the dependence of the cyclotron frequency $\omega$ on the quantum numbers $\left\{n,\,l,\,s\right\}$ of the system can arises in this analogue of the Landau system, whose meaning is that not all values of the cyclotron frequency are allowed in order that bound states solutions can be achieved.

Note that in systems where there is the spontaneous breaking of Lorentz symmetry, the reference frame of the system is the laboratory frame. Recently, rotating effects due to the Earth rotation have been discussed \cite{extra4}. On the other hand, several works \cite{extra3,ens,extra1,extra2,ea,ea2,ea3,col,belich2,lin3,belich,belich1,belich3} have investigated the effects of the violation of the Lorentz symmetry in the laboratory frame by neglecting any effect that stems from the Earth rotation. In the present work, we follow this line of research given in Refs. \cite{extra3,ens,extra1,extra2,ea,ea2,ea3,col,belich2,lin3,belich,belich1,belich3} and neglect any effect from rotation, therefore, the background field is invariant under particle transformations. However, in order to achieve a more realistic measure in a laboratory placed at the surface of the Earth, we should take into account the movement of the earth in relation to a nonrotating frame relative to Sun's center of mass. This kind of measurement was discussed in Ref. \cite{extra4}, where it is shown that the background field becomes a time-dependent term and this should be taken into account when a measurement would be perform.

Recently, an interesting way of estimating a bound for the parameters of the Lorentz symmetry violation with the presence of crossed electric and magnetic fields was proposed in Ref. [17] based on the confinement of a neutral particle to a two-dimensional quantum ring. In the present case, it is very hard to estimate a bound for the parameters of the Lorentz symmetry violation because the magnetic field is considered to be produced by magnetic charges.  
Despite the difficulty of detecting Lorentz symmetry breaking effects and compare with any experimental data, this study opens new discussions of investigating the effects of Lorentz symmetry breaking effects at low energies. Note that we have explored a background of the Lorentz symmetry violation determined by a time-like 4-vector. Other interesting discussions can arise from establishing a background of the violation of the Lorentz symmetry through a space-like 4-vector. Further, a general background of the Lorentz symmetry violation can be described by a fixed 4-vector field that possesses spatial and time components, which means that we would have the spontaneous violation of the $SO(1,3)$ group. This general background brings a more general field settings in which the neutral particle can be submitted. In order to set a background of measurements, it is necessary to choose the better setting according to the type of measurement. We hope to bring this discussion in the near future.

\acknowledgments{We would like to thank CNPq (Conselho Nacional de Desenvolvimento Cient\'ifico e Tecnol\'ogico - Brazil) for financial support.}

\end{document}